\documentstyle[psfig]{mn}

\newif\ifAMStwofonts

\def\etal{et al. }
\def\kcorr{$k$-correction }

\def\hband{$H$-band }
\def\iband{$I$-band }
\def\vband{$V$-band }

\ifoldfss
  \ifCUPmtlplainloaded \else
    \NewTextAlphabet{textbfit} {cmbxti10} {}
    \NewTextAlphabet{textbfss} {cmssbx10} {}
    \NewMathAlphabet{mathbfit} {cmbxti10} {} 
    \NewMathAlphabet{mathbfss} {cmssbx10} {} 
  \fi
  \ifAMStwofonts
    \ifCUPmtlplainloaded \else
      \NewSymbolFont{upmath} {eurm10}
      \NewSymbolFont{AMSa} {msam10}
      \NewMathSymbol{\upi}     {0}{upmath}{19}
      \NewMathSymbol{\umu}     {0}{upmath}{16}
      \NewMathSymbol{\upartial}{0}{upmath}{40}
      \NewMathSymbol{\leqslant}{3}{AMSa}{36}
      \NewMathSymbol{\geqslant}{3}{AMSa}{3E}

    \fi
  \fi
\fi 

\ifnfssone
  \newmathalphabet{\mathit}
  \addtoversion{normal}{\mathit}{cmr}{m}{it}
  \addtoversion{bold}{\mathit}{cmr}{bx}{it}
  \newmathalphabet{\mathbfit} 
  \addtoversion{normal}{\mathbfit}{cmr}{bx}{it}
  \addtoversion{bold}{\mathbfit}{cmr}{bx}{it}
  \newmathalphabet{\mathbfss} 
  \addtoversion{normal}{\mathbfss}{cmss}{bx}{n}
  \addtoversion{bold}{\mathbfss}{cmss}{bx}{n}
  \ifAMStwofonts
    \ifCUPmtlplainloaded \else
      \UseAMStwoboldmath
      \makeatletter
      \new@mathgroup\upmath@group
      \define@mathgroup\mv@normal\upmath@group{eur}{m}{n}
      \define@mathgroup\mv@bold\upmath@group{eur}{b}{n}
      \edef\UPM{\hexnumber\upmath@group}
      \new@mathgroup\amsa@group
      \define@mathgroup\mv@normal\amsa@group{msa}{m}{n}
      \define@mathgroup\mv@bold\amsa@group{msa}{m}{n}
      \edef\AMSa{\hexnumber\amsa@group}
      \makeatother
      \mathchardef\upi="0\UPM19
      \mathchardef\umu="0\UPM16
      \mathchardef\upartial="0\UPM40
      \mathchardef\leqslant="3\AMSa36
      \mathchardef\geqslant="3\AMSa3E
    \fi
  \fi
\fi 

\ifnfsstwo
  \DeclareMathAlphabet{\mathbfit}{OT1}{cmr}{bx}{it}
  \SetMathAlphabet\mathbfit{bold}{OT1}{cmr}{bx}{it}
  \DeclareMathAlphabet{\mathbfss}{OT1}{cmss}{bx}{n}
  \SetMathAlphabet\mathbfss{bold}{OT1}{cmss}{bx}{n}
  \ifAMStwofonts
    \ifCUPmtlplainloaded \else
      \DeclareSymbolFont{UPM}{U}{eur}{m}{n}
      \SetSymbolFont{UPM}{bold}{U}{eur}{b}{n}
      \DeclareSymbolFont{AMSa}{U}{msa}{m}{n}
      \DeclareMathSymbol{\upi}{0}{UPM}{"19}
      \DeclareMathSymbol{\umu}{0}{UPM}{"16}
      \DeclareMathSymbol{\upartial}{0}{UPM}{"40}
      \DeclareMathSymbol{\leqslant}{3}{AMSa}{"36}
      \DeclareMathSymbol{\geqslant}{3}{AMSa}{"3E}
    \fi
  \fi
\fi 

\ifCUPmtlplainloaded \else
  \ifAMStwofonts \else 
    \def\upi{\pi}
    \def\umu{\mu}
    \def\upartial{\partial}
  \fi
\fi

\title{Infrared constraints on the dark mass concentration
observed in the cluster Abell 1942}
\author[M. E. Gray et al.]
       {Meghan E. Gray$^1$\thanks{email: {\em meg@ast.cam.ac.uk}}, Richard S. Ellis$^{1,2}$, James
	R. Lewis$^1$,  Richard G. McMahon$^1$ \cr \&
	Andrew E. Firth$^1$\\
        1. Institute of Astronomy, Madingley Road, Cambridge CB3 0HA \\
        2. Astronomy Department, California Institute of Technology,
        Pasadena CA 91125, USA}
\date{Submitted September 2000}

\pagerange{\pageref{firstpage}--\pageref{lastpage}}
\pubyear{2000}

\begin{document}

\maketitle

\label{firstpage}

\begin{abstract}

We present a deep \hband image of the region in the vicinity of the
cluster Abell 1942 containing the puzzling dark matter concentration
detected in an optical weak lensing study by Erben \etal (2000). We
demonstrate that our limiting magnitude, $H=22$, would be sufficient
to detect clusters of appropriate mass out to redshifts comparable
with the mean redshift of the background sources. Despite this, our
infrared image reveals no obvious overdensity of sources at the
location of the lensing mass peak, nor an excess of sources in the
$I-H$ vs. $H$ colour-magnitude diagram. We use this to further constrain
the luminosity and mass-to-light ratio of the putative dark clump as a
function of its redshift.  We find that for spatially-flat
cosmologies, background lensing clusters with reasonable mass-to-light
ratios lying in the redshift range 0$<$z$<$1 are strongly excluded,
leaving open the possibility that the mass concentration is a new type
of truly dark object.
\end{abstract}

\begin{keywords}
gravitational lensing -- galaxies: clusters: individual: Abell 1942 --
infrared: galaxies -- cosmology: dark matter
\end{keywords}

\section{Introduction}

Gravitational lensing offers the most practical route to tracing
the spatial distribution of dark matter and its relationship to
that of the radiating baryonic component. Most progress has been
made in rich galaxy clusters through both detailed studies of
strongly-lensed features of known spectroscopic redshift (e.g.
Kneib et al. 1996) and weak lensing analyses, where elaborate tools
have been developed to convert the measured coherent alignment of
faint background galaxies into the projected cluster mass density
field (Kaiser \& Schneider 1993, Schneider 1996). The resulting lensing
mass distributions appear to agree reasonably well with those
inferred from the optical light over a range of scales (Smail et
al 1995, Bonnet et al. 1994). Although discrepancies have been
reported in the concentration of the inferred dark matter relative
to that of the baryonic component (Miralda-Escud\'e \& Babul
1995), the uncertainties remain significant and there is not yet a
compelling case for distinct distributions of dark and visible
matter (Smail et al. 1996, Allen 1998).

Deep imaging surveys are now beginning to explore the spectrum of
mass concentrations outside rich clusters (Wittmann et al. 2000,
van Waerbeke et al. 2000, Bacon et al. 2000, Kaiser et al. 2000).
These so-called `cosmic shear' field surveys are primarily
concerned with a statistical description of the weak lensing
signal as a function of angular scale. They could also provide
further constraints on the presence or otherwise of concentrations
with high mass-to-light ratios. Although the detection of dark
halos without significant baryonic material would represent a
considerable challenge to standard theories for the origin of
structure, it is important to examine the observational
constraints.

In an important paper, Erben et al. (2000) report the first detection
of a dark matter halo not obviously associated with any light. Using
two wide field optical images of the rich cluster Abell 1942
($z$=0.224), they find a highly significant secondary peak in the
reconstructed mass distribution located about 7 arcmin south of the
cluster centre (in addition to also finding a strong signal
corresponding to the cluster itself). An archival ROSAT/HRI image
shows a 3.2$\sigma$ source close (but not precisely co-located) with
this secondary mass peak. A marginal overdensity of galaxies in their
\iband image is claimed by Erben et al.  The authors are unable to
associate the lensing signal with the X-ray source on the assumption
that both represent a distant projected cluster and conclude that the
mass concentration identified via their lensing study is either a
distant cluster unrelated to the X-ray source, or tantalisingly, a
dark halo of $\simeq$3$\times$10$^{14}$ $M_{\odot}$ located within the
cluster.

The burden of proof for the existence of such a dark clump within
Abell 1942 is high. As Erben et al. discuss, the mass-to-light
ratio inferred depends sensitively on its (unknown) redshift.
Quite reasonable mass-to-light ratios are implied if the clump is
placed at high redshift but the inferred mass then also increases.
The most conservative explanation would place the clump at
$z\simeq$0.8 with $M/L>$450 (for an Einstein-de Sitter cosmology).
Deeper imaging at infrared wavelengths represents one obvious way
to eliminate or confirm this explanation.

In this paper we follow up the optical study of Erben et al. (2000)
with a deep $H$-band (1.6 $\mu$m) image analysis. As the infrared
\kcorr is considerably smaller and more uniform across the
Hubble sequence of galaxy types than is the case at optical
wavelengths, deep infrared imaging is a powerful tool for
exploring the existence or otherwise of all galaxy populations at
high redshifts. Our goal is to extend the analysis of Erben et al.
and to attempt to locate the origin of this puzzling lensing
signal.

Our paper is structured as follows. In $\S$2 we discuss the unique
advantages of using an infrared imager to constrain the properties of
the proposed dark clump by demonstrating that, at achievable limiting
$H$ magnitudes, we can reasonably expect to see most conventional
clusters to the mean redshift of the background population used by
Erben et al. (2000). In $\S$3 we describe our deep \hband observations
of Abell 1942 and discuss their reduction and suitability in terms of
our achieved limiting magnitude. $\S$4 presents our analysis of the
likely \hband light associated with both the dark matter clump and the
nearby X-ray source. No convincing excess is seen in either case and
we convert the observed flux limits into new constraints on the
mass-to-light ratio of the putative clump.  We present our conclusions
in $\S$5.

\section{Strategic Issues}

Erben et al. (2000) imaged Abell 1942 in both the $V$ and $I$
photometric bands using different imaging cameras on the 3.5m
Canada-France-Hawaii Telescope. The dark clump was inferred from the
weak shear signals observed in both datasets. Although their 4.5 hour
\vband image provides the most significant detection of the dark
clump, because of the strong \kcorr for early type galaxies beyond
$z\simeq$0.5, the 3 hour \iband image provides the most interesting
constraints on the presence of a distant lensing cluster.

\begin{figure}
\centerline{\psfig{figure=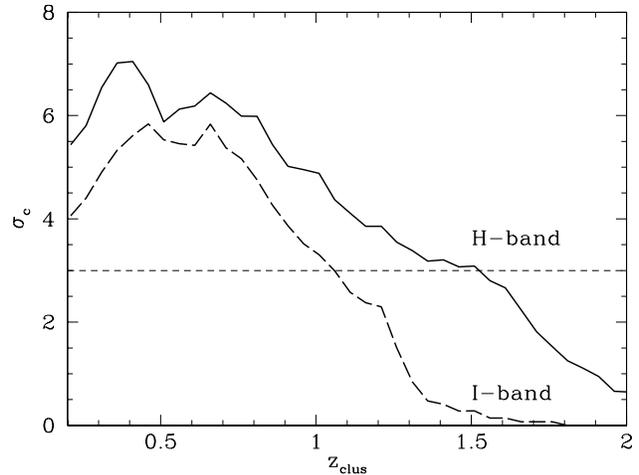,width=0.5\textwidth,angle=270}}
\caption{The observed contrast of a richness class III cluster against
the background field as a function of redshift for the $I$ and $H$
photometric bands, assuming magnitude limits of 24.5 and 22
respectively. The simulation is based on the $I-H$ colour-selected
luminosity function of Abell 2219 (Gray et al. 2000) scaled to the
angular size of the putative dark clump in Abell 1942 (60 arcsec)
using in a spatially flat cosmology with $\Omega_M$=0.3,
$\Omega_{\Lambda}$=0.7. The results are essentially identical for an
Einstein-de Sitter universe.}
\label{fig-contrast}
\end{figure}

Erben \etal claim the mean redshift of the background population at
their limit of $I\simeq$24.5 is $\simeq$0.7-0.8 (although no
spectroscopic data yet exists to such a faint limit and a higher mean
redshift cannot be ruled out). A key strategic question for this paper
is, therefore, the depth necessary at near-infrared wavelengths
(specifically in the \hband in which our observations are conducted),
to better detect background clusters {\em up to the redshift limit
appropriate for the \iband image, say $z\simeq$1.}

Using the $I-H$ colour $k$-correction, $k_{IH}(z)$, for spheroidal
galaxies taken from the analysis of Menanteau et al. (1999), a
cluster elliptical with $I$=24.5 at the mean redshift of the
background population would be expected to have $H$=21.7. However,
there is a further advantage of conducting cluster searches in the
infrared in that the contrast of cluster members at faint limits
against the field population is much greater than in the optical
because of the smaller optical \kcorr for the bulk of the
intervening field population.

We can define the observed contrast of a cluster \mbox{$\sigma_c = N_{\rm
cl}/\sigma_f$} (following Couch \etal 1991), where $N_{\rm cl}$ is the
field-subtracted number of cluster galaxies brighter than the limiting
magnitude of the survey within some physical aperture, and $\sigma_f$
is the rms of the background field counts on the scale of that
aperture.  We will investigate the effects of the \kcorr in the $I$-
and $H$-bands on the visibility of a rich cluster with redshift, using
previously presented data for Abell 2219 (Gray \etal 2000).

Abell 2219 is a richness class III cluster at $z=0.22$, a similar
richness and redshift to Abell 1942.  We determine $N_{\rm cl}$ by
counting galaxies within an aperture of radius 60 arcsec
(corresponding to a physical radius of $125 h^{-1}$ kpc at this
redshift) using the colour selection of Gray \etal (2000).  By
comparing this number with that found with the same aperture and
colour selection applied to offset fields we determine that roughly
20\% of these are field galaxies, and correct $N_{\rm cl}$ by this
factor.  Using $k$-corrections of Menanteau \etal (1999) we may
determine what $H$- and $I$-band magnitudes these cluster galaxies
would have if the cluster were shifted to higher redshift, and what
fraction of those galaxies would then be brighter than the limiting
magnitudes of $H=22$ and $I=24.5$.  At each redshift we measure
$\sigma_f$ of the field counts in an aperture corresponding to the
angular size of the physical aperture at that redshift, and calculate
the contrast, $\sigma_c$.

Fig.~\ref{fig-contrast} shows the results of this calculation.  Abell
2219 itself at $z=0.22$ yields a $4\sigma$ detection in the \iband and
a $5\sigma$ detection in the $H$-band.  To the $3\sigma$ level we would
be able to see such a cluster to $z=0.9$ in $I$ and $z=1.4$ in
$H$. This demonstrates the advantage of searching for clusters in the
infrared: for a given cluster, the contrast is greater and remains
greater to much higher redshift due to the advantageous effects of the
\kcorr and reduced numbers of field galaxies.  If the unexplained mass
detection of Erben \etal (2000) were the result of a high-redshift
cluster similar to Abell 1942 or Abell 2219, then for a limiting
magnitude $H=22$ we would expect to be able to make a $\sigma>4$
detection in the \hband to the redshift limit they present for such a
cluster.

For such a magnitude limit we may also determine how poor a cluster we
could detect.  For a cluster at $z=1$ Fig.~\ref{fig-contrast} shows
that a richness class III cluster has $\sigma_c \simeq 5$.  We would
therefore detect a cluster at that redshift with 3/5 as many galaxies
at the $3 \sigma$ level.  Using the relations between numbers,
richness, and mass of van Kampen \& Katgert (1997) this means that to
$H=22$ we would be able to detect a cluster with roughly 80\% the mass
of Abell 2219, or a richness class II cluster.

\section{Observations} \label{sec-obs}

\begin{table}
\centering
\begin{tabular}{lrr}
\hline
\multicolumn{1}{c}{date} & \multicolumn{1}{c}{exposures} & \multicolumn{1}{c}{total exposure time}\\
& & \multicolumn{1}{c}{(seconds)} \\
\hline
2000 Feb 17     & 116   & 5220\\
2000 Feb 18 	&  21   &  945\\
2000 Feb 19     &  78   & 3510\\
2000 Mar 19 	&  27   & 1215\\
\hline
\end{tabular}
\caption{Summary of CIRSI \hband observations of the Abell 1942/dark
lump region.  The combined total exposure time is 10890 seconds.}
\label{tab-obs}
\end{table}

\begin{figure} 
\centerline{\psfig{figure=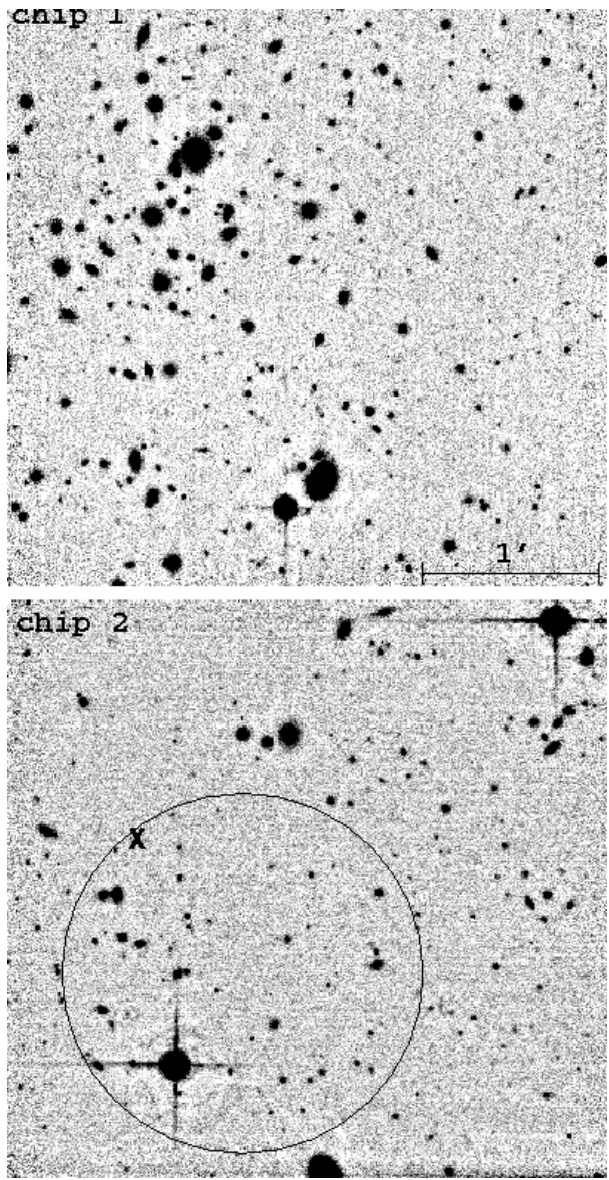,width=0.45\textwidth,angle=0}}
\caption{Two of the four deep CIRSI \hband $3.4 \times 3.4$ arcmin
images.  {\em Top}: core of Abell 1942 (chip 1).  {\em Bottom}: dark
lump region (chip 2; actually 7 arcmin south of chip 1).  The position
of the X-ray source discussed in Erben \etal is marked by an X; an
aperture of radius 1 arcmin is shown centred over the dark lump
region.  The two remaining chip images (chips 3 and 4) are not shown,
but they each image offset fields which lie $\sim$7 arcmin east of the
fields shown.  North is up, east is to the left. }
\label{fig-image}
\end{figure}

Infrared imaging data for the rich cluster Abell 1942 ($z=0.224$)
were taken over four nights in February and March 2000 using the
Cambridge Infrared Survey Instrument (CIRSI, Beckett \etal 1998)
on the 2.5m du Pont telescope at Las Campanas, Chile. CIRSI is a
wide-field infrared imager composed of four 1K $\times$ 1K
Rockwell Hawaii HgCdTe detectors, with a gap of approximately one
detector width between each array.  On the Dupont telescope the
pixel scale is 0.2 arcsec/pixel, corresponding to a field of view
of $3.4 \times 3.4$ arcmin for each detector.

The instrument pointing was arranged so that one chip imaged
the core of Abell 1942, one chip imaged the dark lump region
as indicated by the mass maps of Erben \etal (2000), and the
remaining two chips covered adjacent blank sky regions.  This
configuration ensured that the infrared data would overlap the
existing CFHT UH8K \iband image (which has a similar pixel size of
0.24 arcsec/pixel) whilst providing essential background field count
data.

The deep CIRSI \hband observations were made using 45 second
exposures with a nine-point dither pattern.  Typically three or
four exposures were taken at each dither position, followed by an
offset of 9 or 15 arcsec.  Table~\ref{tab-obs} summarizes the
observations.  

The CIRSI data were reduced using the pipeline developed by one of us
(JRL) for the Las Campanas Infrared Survey (Marzke \etal 1999, Firth
2000).  Flat field frames were created from dome flats and bad pixel
masks constructed from the data. The images at each dither position
were coadded and flat fielded, followed by a first-pass sky
subtraction.  Object masks were constructed for the sky-subtracted
images and these masks were used to perform a second-pass running sky
subtraction on the original images.  A reset anomaly present in the
data was corrected by median filtering the rows and columns of each
quadrant of each image.  The offsets between images due to the
dithering pattern were calculated, and a tangent plane world
coordinate system (WCS) was fitted.  A final weighted image was
constructed for each chip using the WCS for each exposure.  The
regions of low S/N where the overlap was not complete were discarded,
and the resulting images measured $3.4 \times 3.4$ arcmin per chip.
The median seeing for the final coadded images was 0.7$\arcsec$.

Standard stars were chosen from the near-infrared catalogue of Persson
\etal 1998.  Dither sequences were calibrated using the standard star
observations from the relevant night.  A set of secondary standards
within the sub-images were then measured and used to calculate the
zeropoint of the final weighted image.  Fig.~\ref{fig-image} shows
two of the four resulting chip images: the cluster core and the `dark
lump' region.

Object catalogues were created for each chip image using SExtractor
2.0 (Bertin \& Arnouts 1996) with a detection threshold of 1.5$\sigma$
above the noise level and requiring an object to have at least 6
connected pixels. Isophotal magnitudes were measured separately for
the CIRSI \hband and the UH8K \iband image (described in Erben \etal
2000). SExtractor was then run in double image mode to measure an
$I-H$ colour using the \iband isophot (when registered to the \hband
image).  Additional catalogues were created encompassing those objects
seen in only one of the bands.

\begin{figure}
\centerline{\psfig{figure=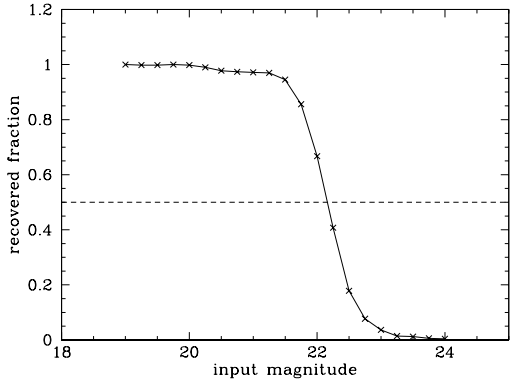,width=0.4\textwidth,angle=270}}
\caption{Completeness study for the CIRSI images of Abell 1942 based on artificial
image tests indicating a 50\% completeness at $H$=22.}
\label{fig-completeness}
\end{figure}

The turnover in the number counts reveals magnitude limits of $H=22.5$
and $I=24.5$ (the latter in agreement with the results quoted by Erben
\etal) for the two images.  To further test the magnitude limits of
the \hband data we performed completeness tests.  Using the {\tt
artdata} package in IRAF we added galaxies of known magnitude to the
data and recovered them using SExtractor as above.  The simulations
were performed ten times, with 50 galaxies being added in random
locations each time.  Fig.~\ref{fig-completeness} shows that a 50\%
completeness level is achieved at a limiting magnitude of $H=22$.

\section{Analysis}

\subsection{Search for a high-redshift cluster}

We begin our analysis of the dark clump region by adopting the
hypothesis that the mass concentration arises from a background
cluster whose contrast was not sufficiently large in the optical
images available to Erben \etal (c.f. Fig.~\ref{fig-contrast}).  We
will first examine this possibility in terms of the \mbox{$I-H$}
colours of sources in the region of the dark lump, later examining the
possibility of a cluster recognisable via images found only via their
detection in the CIRSI image.

\subsubsection{Colour-based search}

The key ingredients necessary for recognising a distant cluster are a
faint limiting magnitude and contrast against the background field.
Conceivably, a distant cluster can be rendered visible within Erben et
al.'s $I$-band image by combining with our $H$-band image and searching
for the signature of an appropriately red colour-magnitude relation of
its early-type galaxies so that its contrast is increased.  Early-type
galaxies in clusters have been shown to obey a linear relation on the
colour-magnitude diagram with extremely small scatter (Bower, Lucey \&
Ellis 1992a,b). This relation has been exploited to detect and confirm
clusters via their red sequence (Gladders \& Yee 2000, da Costa \etal
1999).

\begin{figure*}
\centerline{\psfig{figure=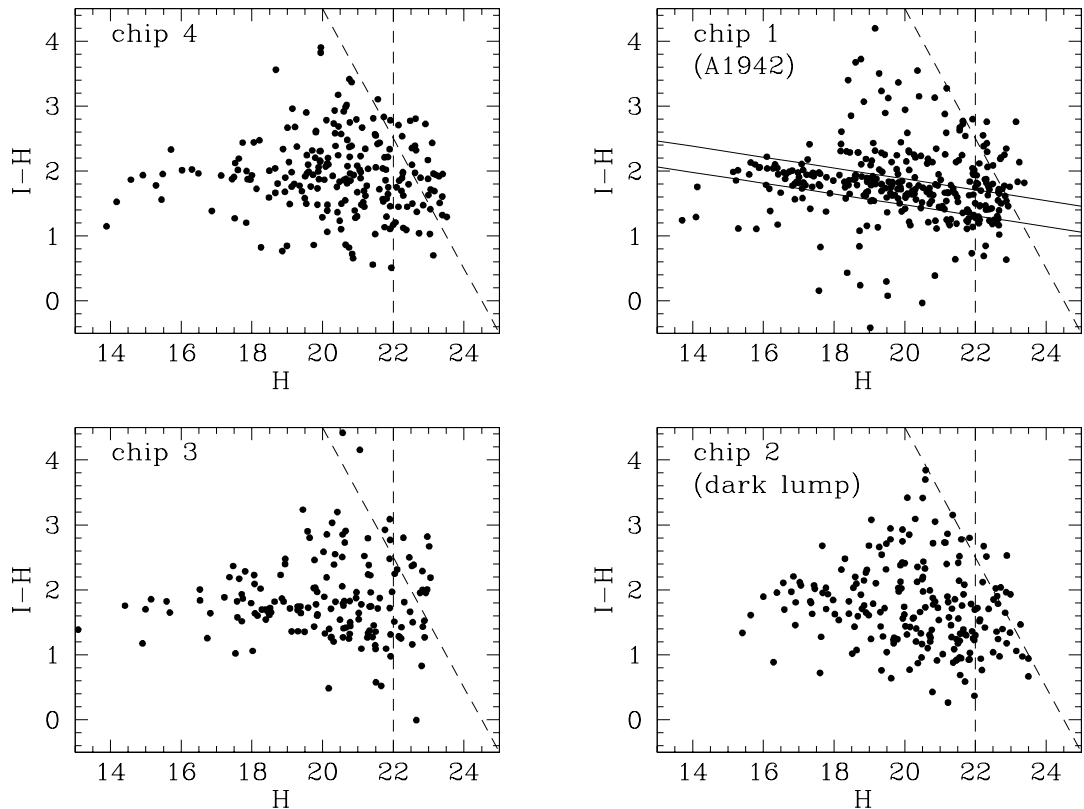,width=0.9\textwidth,angle=270}}
\caption{Colour-magnitude diagrams for the four CIRSI chips from the
combined $I$- and \hband catalogues.  The dashed lines indicate the
$I<24.5$ and $H<22$ magnitude limits of the samples.  No obvious
early-type sequence is seen for the dark lump region (chip 2), in
contrast to the $I-H$ selected Abell 1942 sequence in chip 1
(delineated by the parallel lines).}
\label{fig-colmag}
\end{figure*}

Fig.~\ref{fig-colmag} shows the colour-magnitude diagrams for the four
chips based on the matched $I,H$ catalogues.  In the case of chip 1
containing the Abell 1942 cluster core, the sequence of $z$=0.22
early-type galaxies is clearly visible.  However no sequence with
redder colours can be detected in the chip 2 containing the dark
matter lump (or, in fact, in any other chip).  We quantified this
result by searching for an excess contrast in chip 2 c.f. the mean of
that found in the other chips according to colour slices selected with
an appropriate colour-magnitude slope.

The $I-H$ cluster sequence bounded by the parallel lines in
Fig.~\ref{fig-colmag} is described by the relation
\[
\left| (I-H) - (3.35 - 0.0836H) < 0.2 \right| ,
\]
which agrees well with the colour-magnitude relation measured in Abell
2219 using the same filters (Gray \etal 2000).  We use this gradient,
$m=-0.0836$, to divide the colour-magnitude diagram for each of the
four CIRSI chips into colour bins.  We choose a width of 0.2 mag for
the colour slices, and overlap successive bins to allow for any
potential cluster galaxies that may span more than one bin.  Comparing
the excess numbers in each bin with the $\sqrt N$ errors derived from
the average for that bin for chips 2,3, and 4 (those chips not
containing the core of Abell 1942) we calculate a contrast,
$\sigma_{\rm col}$, for each colour slice and each chip.  These
results are summarized in Table~\ref{tab-colslice}.

\begin{table}
\centering
\begin{tabular}{rrrrr|rr}
\hline
\multicolumn{1}{c}{$I-H$} & 
\multicolumn{1}{c}{$\sigma_{\rm col}$} &
\multicolumn{1}{c}{$\sigma_{\rm col}$}&
\multicolumn{1}{c}{$\sigma_{\rm col}$} &
\multicolumn{1}{c}{$\sigma_{\rm col}$} &
\multicolumn{1}{c}{$\sigma_{\rm sim}$} & 
\multicolumn{1}{c}{$z$} \\
\multicolumn{1}{c}{\scriptsize $(H=19)$} & \multicolumn{1}{c}{\scriptsize chip
1}& \multicolumn{1}{c}{\scriptsize chip 2}& \multicolumn{1}{c}{\scriptsize chip
3}& \multicolumn{1}{c}{\scriptsize chip 4} & \multicolumn{1}{c}{\scriptsize (A2219)} & \\
\hline
   1.012  &  0.45  &  1.79  &  1.34  &  0.45  &   &  \\
   1.112  &  0.65  &  2.84  &  2.19  &  0.65  &   &  \\
   1.212  &  0.33  &  2.67  &  2.00  &  0.67  &   &  \\
   1.312  &  0.27  &  1.07  &  0.80  &  1.87  &   &  \\
   1.412  &  0.31  &  0.31  &  1.23  &  1.54  &   &  \\
   1.512  &  1.75  &  1.09  &  0.22  &  1.31  &   &  \\
   1.612  &  4.90  &  0.82  &  0.41  &  1.22  &   &  \\
   1.712  &  {\bf 8.16}  &  0.53  &  0.07  &  0.46 &    &\\
   1.812  &  {\bf 13.13} &  0.29  &  0.94  &  1.23 &     4.60 &     0.30\\
   1.912  &  {\bf 7.60}  &  0.00  &  0.89  &  0.89 &     5.07 &     0.35\\
   2.012  &  0.29  &  0.29  &  0.51  &  0.79 &     5.52 &     0.40\\
   2.112  &  0.08  &  0.08  &  0.82  &  0.91 &     5.67 &     0.45\\
   2.212  &  1.71  &  0.45  &  1.26  &  1.71 &     5.63 &     0.48\\
   2.312  &  {\bf 3.46}  &  0.87  &  2.02  &  2.89 &     5.63 &     0.53\\
   2.412  &  1.53  &  0.11  &  1.42  &  1.53 &     5.83 &     0.59\\
   2.512  &  0.60  &  0.84  &  0.60  &  0.24 &     5.04 &     0.79\\
   2.612  &  1.92  &  0.57  &  0.91  &  1.47 &     5.60 &     0.82\\
   2.712  &  1.23  &  0.86  &  0.86  &  1.72 &     5.17 &     0.93\\
   2.812  &  0.28  &  0.56  &  1.12  &  0.56 &     5.20 &     0.98\\
   2.912  &  0.86  &  0.62  &  0.49  &  0.12 &     5.87 &     1.00\\
   3.012  &  0.14  &  0.14  &  0.28  &  0.14 &     5.99 &     1.06\\
   3.112  &  0.17  &  0.17  &  0.87  &  0.70 &     6.04 &     1.12\\
   3.212  &  1.09  &  0.22  &  0.22  &  0.44 &     6.32 &     1.18\\
\hline
\end{tabular}
\caption{Significance of any excess of galaxies in each colour slice
for each of the four CIRSI chips.  The $I-H$ colour in column 1 refers
to the central colour of the bin at $H=19$, and the gradient used is
that measured for Abell 1942, $m=-0.0865$.  Abell 1942 itself is
detected as a 13$\sigma$ excess for $I-H=1.812$, while no $\sigma_{\rm
col} > 3$ colour sequence (in bold) is seen for any bin in chips 2, 3,
or 4.  For comparison a similar calculation was performed for the
redshifted A2219 simulations described in $\S2$.  Column 5 gives the
significance for such a richness class III cluster falling within the
given colour bin, and column 6 lists the associated cluster redshift
which yields that colour.  As in Fig.~\ref{fig-contrast} we find that
if the mass detection of Erben \etal were the result of a rich cluster
similar to Abell 2219, it would be significantly detected as an excess
within the colour-magnitude diagram out to $z=3$.}
\label{tab-colslice}
\end{table}

Abell 1942 is detected in the $I-H=1.812$ and neighbouring bins for
chip 1 with a peak significance of $\sigma_{\rm col}=13.13$.  However,
the remaining three chips (in particular chip 2, containing the
putative dark lump) show no colour slices for which $\sigma_{\rm
col}>3$.  We do, however, see a $\sigma_{\rm col}=3.46$ detection for
the $I-H=2.312$ bin in chip 1.  This marginal detection corresponds to
the sequence visible with colours $\sim 0.4$ mag redder than the Abell
1942 sequence in Fig.~\ref{fig-colmag}.  These colours are consistent
with those of early-type galaxies at $z=0.5$.  However, when the
positions of the 25 galaxies in this bin are plotted on the sky, they
show no obvious clustering and are spread nearly evenly across the
four quadrants of chip 1. They are therefore not likely to be in
physical association.  Similarly, a slight overdensity is seen in the
$I-H = 1.112$ and $1.212$ bins in chip 2 but these galaxies show no angular
clustering and are not co-located with the dark lump region.

\begin{figure*}
\centerline{
\psfig{figure=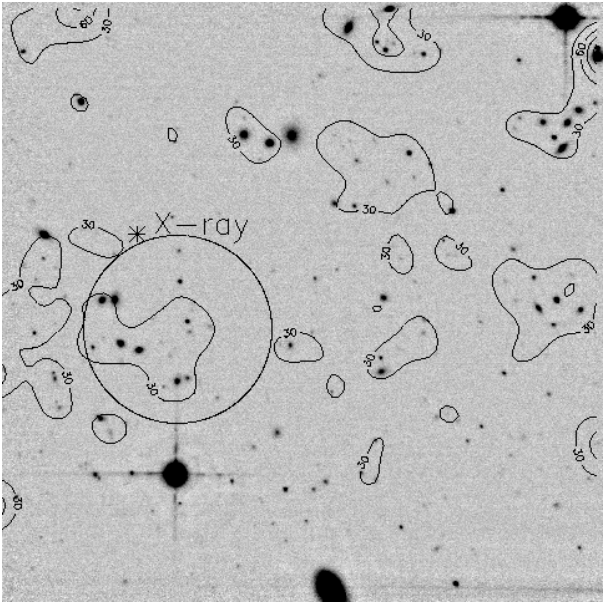,width=0.45\textwidth,angle=0}
\psfig{figure=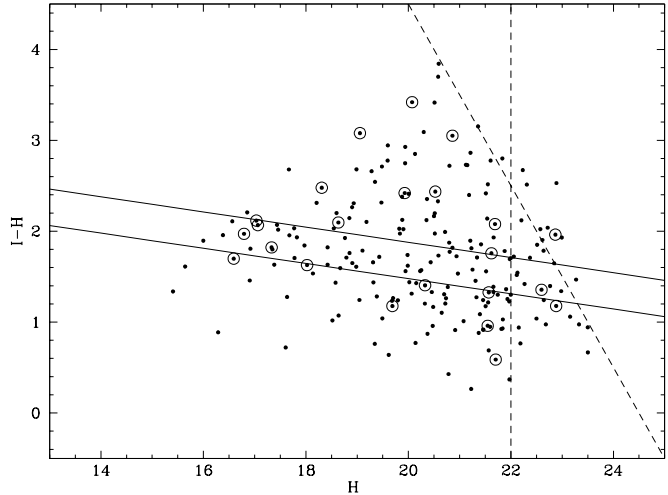,width=0.55\textwidth,angle=270}
}
\caption{{\em Left}: Contours of surface number density of galaxies
(arcmin$^{-2}$) in the matched $I$- and $H$-band catalogue for the
dark lump region (smoothed with a Gaussian of width $\sigma=6$
arcsec).  The 30 arcmin radius aperture shown here encircles the
slight overdensity of galaxies noted in Erben et al. lying near to the
position of the X-ray source (marked with a star). {\em Right}: open
symbols show the location of the galaxies within this aperture on the
colour-magnitude diagram for entire chip.  Clearly these galaxies are
not physically associated within a cluster.  The brightest galaxies
have colours similar to those for the Abell 1942 cluster galaxies
(delineated as in Fig.~\ref{fig-colmag} by two parallel lines).}
\label{fig-overdensity}
\end{figure*}

Erben \etal reported a slight overdensity of galaxies in their
$I$-band image approximately 60 arcsec from the location of the mass
peak and close to the X-ray source.  We also detect this marginal
concentration in the matched $I$- and $H$-band catalogues
(Figure~\ref{fig-overdensity}).  However, when we examine the location
of those galaxies in this excess region on the colour-magnitude
relation, no obvious trend is found.  Even allowing for field
contamination, at most 2-3 galaxies have $I-H$ colours appropriate for
a high redshift cluster and the colour variation from galaxy to galaxy
is considerable.

\subsubsection{Population of H-only objects?}

Next, we turn our attention to the faintest $H$-band objects in the CIRSI
image {\em including those sufficiently faint that Erben et al. did not
detect them in their $I$-band image}. As we showed in $\S$2, the reduced
$k$-correction in the $H$-band makes it possible that a distant cluster
could be seen in the CIRSI image but not in the $I$-band data (c.f. Figure
1).

In this case we utilised a SExtractor catalogue based only on the
$H$-band image, foregoing the formal magnitude limit on the assumption
that the magnitude-dependent selection function is positionally
invariant across the detectors (Gray et al. 2000). Figure 6 shows the
location of all the sources that consist of at least 5 connected
pixels whose surface brightness lies 1$\sigma$ above the mean
background.  Again, the number of sources within an aperture centred
on the location of the dark lump deviates by less that $1\sigma$ from
the mean value determined for 10 randomly-placed apertures on the same
chip.  This statement is true both for sources to $H=22$ and to the
catalogue limit, and for apertures of 18 and 36 arcsec radii
(corresponding to the $1\sigma$ and $2\sigma$ dispersion for the
centroid of the Erben et al. mass peak).  As shown previously in
$\S2$, we expect a significantly stronger signal in the case of even a
richness class II cluster at $z\simeq 1$.

In conclusion, neither the colour-based nor the $H$-band search has turned
up a convincing case for any physical association of background sources in
the region of the mass peak. Most importantly, the colours of sources in
this region show a wide variation and, to a faint $H$-limit, no excess is
visible.

\begin{figure}
\centerline{\psfig{figure=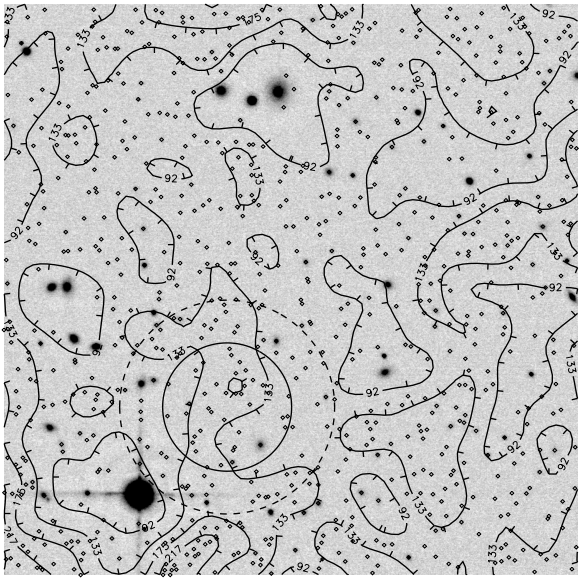,width=0.45\textwidth,angle=0}}
\caption{Contours of surface number density (arcmin$^{-2}$) for the
deep \hband catalogue, smoothed with a Gaussian of width $\sigma=6$
arcsec.  Sticks point in the direction of decreasing surface density.
No significant excess of infrared objects is found within an aperture
of 18 arcsec (solid) or 36 arcsec (dashed) radius centred on the
location of the mass concentration (as determined from the Erben et
al. mass reconstruction).}
\label{fig-hcontours}
\end{figure}

\subsection{A Dark Mass Concentration} \label{sec-ml}

Having discounted the possibility of a conventional background cluster of
galaxies, we next attempt to place limits on the $H$-band luminosity
associated with the mass peak and hence the mass-to-light ratio of the
structure which will depends on its (unknown) redshift.

We first consider the area around the X-ray source located at
$\alpha=14^{\rm h} 38^{\rm m} 22.8^{\rm s}$, $\delta=3^{\circ}
33\arcmin 11\arcsec$.  No obvious infrared source is found at this
location.  We sum the flux of all the non-stellar objects within a
radius of 20 arcsec that are contained in the \hband photometric
catalogue described in Section~\ref{sec-obs} to $H=22$.  Comparing to
the flux contained in 26 control apertures distributed on the three
non-cluster chips we find the flux in the aperture containing the
X-ray source is only 1.4$\sigma$ above the mean.

Similarly, we examine the \hband flux of objects within 50 arcsec of
the location of the mass concentration.  Again, there is no
significant excess of \hband flux compared to that within control
apertures.

\begin{figure}
\centerline{\psfig{figure=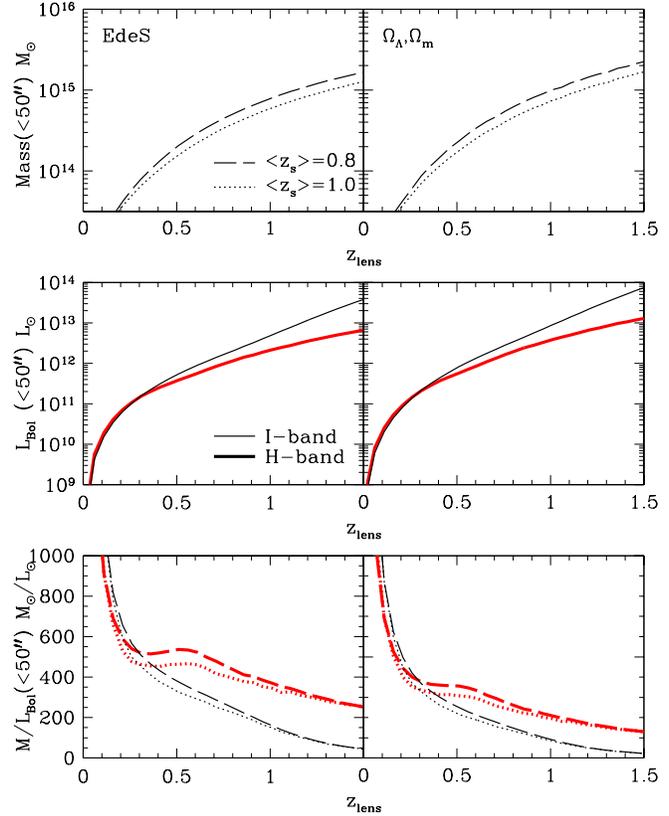,width=0.5\textwidth,angle=0}}
\caption{Lensing mass, bolometric luminosity, and $M/L_{Bol}$ as a
function of lens redshift for the dark clump within a 50 arcsec
aperture.  The left hand panels assume an Einstein-de Sitter
cosmology; the right-hand panels assume $\Omega_\Lambda=0.7,
\Omega_M=0.3$, both with $H_0=50$km s$^{-1}$ Mpc$^{-1}$).  {\em Top}:
lensing mass within the observed aperture for $\left< z_s \right> =
0.8$ and $\left<z_s\right> = 1.0$, assuming the redshift distribution
of Erben et al.  {\em Middle}: the upper limits on the bolometric
luminosity as a function of lens redshift.  The thin and thick lines
show the properties derived from the $I$-band image of Erben \etal and
the $H$-band data presented in this paper, respectively.  The smaller
$k$-correction in the infrared results in tighter constraints at high
redshift.  {\em Bottom}: The {\em lower limits} on the mass-to-light
ratio within the aperture for the combinations of the two source
redshifts and two photometric bands.  The addition of the infrared
data provides stronger constraints on the $M/L_{Bol}$ of the lens and
show it to be significantly darker than a conventional cluster to high
redshift.}
\label{fig-ml}
\end{figure}

Finally, we consider the constraints placed on the mass-to-light ratio
of the mass concentration.  The measured \hband flux within the
50\arcsec aperture is not significantly in excess of that found in a
sample of control apertures placed randomly on chips 2,3, and 4
(avoiding Abell 1942 on chip 1).  We use the $1\sigma$ fluctuations in
the \hband flux in the control apertures to place an upper limit on
the excess luminosity, which will be dependent on the cosmology and
the unknown redshift of the dark clump.  The total apparent flux is
transformed into a total \hband luminosity, using the Menanteau \etal
$k$-corrections for early-type galaxies.  This measurement of $L_H$ is
then transformed into the bolometric luminosity $L_{bol}$ using
bolometric corrections for an old stellar population from the same
models.  This curve, which varies with the unknown redshift of the
clump, is shown in Fig.~\ref{fig-ml} and allows comparison with the
curves derived in a similar fashion from the $I$-band data.  As these
are upper limits on the luminosity of the clump, the $H$-band data
provides tighter constraints on the luminosity at high redshift.

The weak lensing measurements of Erben \etal provide an estimate of
the lensing mass within the aperture, assuming a redshift distribution
$\propto z^2 \exp(-(z/z_0)^{1.5})$ for the source galaxies, with
$\left< z_s \right>=0.8$ and $\left< z_s \right>=1.0$ (where $\left<
z_s \right>\simeq 1.5z_0$).  Combining these lensing mass estimates
with the upper limits on the bolometric luminosity due to the cluster
galaxies, we obtain the lower limits on the $M/L_{Bol}$ shown in the
bottom panel of Fig.~\ref{fig-ml}.

\section{Discussion}

From our infrared study we have discounted the possibility that the
mass concentration detected by Erben \etal is the result of a hitherto
undetected massive high redshift cluster of galaxies.  The resulting
velocity dispersion, $\sigma_v$, implied by the measured tangential
shear is in excess of 2000 km s$^{-1}$ for a singular isothermal
sphere model with $\left<z_{\rm lens}\right>=1.0$.  This is
considerably higher than any currently measured rich cluster, and we
have shown in Fig.~\ref{fig-contrast} that we would expect to make a
significant detection of such a cluster to the magnitude limit of our
infrared study.  Both searches for an excess in the $I-H$
colour-magnitude diagram and a spatial overdensity of sources in the
deep infrared data reveal no such features.  Furthermore, a massive
cluster at higher redshift (beyond the magnitude limit of our study)
is ruled out by the lensing constraint provided by the finite redshift
of the background sources (for which a reasonable limit is $\left< z_s
\right>=1.0$): the lensing object cannot be more distant than the
sources being lensed.

The lower limits on the mass-to-light ratio inferred in
$\S$\ref{sec-ml} confirm the darkness of the mass concentration, and
the infrared data are shown to provide stronger constraints on these
limits at high redshift.  Changing the redshift distribution of the
sources used in the lensing reconstruction from $\left< z_s \right> =
0.8$ to $\left< z_s \right> = 1.0$ serves to lower the lensing mass
and hence the resulting $M/L$, but only slightly.  Removing the
discrepancy would require the sources used in the lensing mass
reconstruction to be placed at significantly higher redshift.  We find
$M/L_{Bol} > 500$ for $0.2<z_{lens}<1$ derived from the $H$-band
measurements (slightly lower in a $\Lambda$-cosmology).  

Mass-to-light ratios for clusters presented in the literature tend to
be calculated in blue or visible light, which makes direct comparisons
difficult (although if bolometric corrections similar to those applied
in the calculations above are followed, transforming $L_V$ to $L_{\rm
Bol}$ will {\em lower} the resulting $M/L_{Bol}$).  Hradecky \etal
(2000) present $M/L_V$ for a sample of eight low-redshift relaxed
clusters with masses derived from the X-ray properties.  They find a
median $M/L_V \sim 100 h_{50}$ in solar units.
Other studies using weak lensing techniques have found moderately
higher cluster mass-to-light ratios, e.g. $M/L_V=270h$ for MS1137
($z=0.783$) and $M/L_V=190h$ for RXJ1716 ($z=0.813$) (Clowe \etal 1998),
$M/L_V=300 \pm 100 h$ for Abell 2163 (Squires \etal 1997), ranging as
high as $M/L_R=640\pm 150$ for the exceptionally dark cluster MS 1224
at $z=0.3$ (Fischer \etal 1999, Fahlman \etal 1994).  Clearly, the
lower limits we find for the mass concentration place it firmly at the
`dark' end of the $M/L$ ratio scale for conventional clusters.

The reality of the lensing signal, which has been verified by Erben et
al. in two separate datasets using two different cameras, is not in
question.  The absence of any overdensity of infrared sources or light
makes the mass concentration increasingly enigmatic.  Having ruled out
a high-redshift cluster we are left to consider the possibility of a
new type of object comprising a truly dark halo with an extremely high
mass-to-light ratio.  

Alternative explanations would involve chance alignments of the
background galaxies at the location of the lensing signal (shown by
Erben et al. to have a probability $< 10^{-4} - 10^{-6}$) or would
require placing the source galaxies at much higher redshift than
previously expected.  A more complex lens configuration involving
projection effects along the line of sight (Metzler, White, \& Loken
2000) could also be invoked.  Finally, recent studies regarding the
intrinsic correlations of galaxy shapes (e.g. Heavens, Refregier, \&
Heymans 2000) and the resulting implications for weak lensing studies
must also be considered, as such intrinsic alignments could greatly
increase the probability of uncovering a false mass peak.

Further studies to obtain photometric redshifts of the background
sources, space-based high resolution data for improved lensing
reconstructions, and deeper X-ray or SZ studies of the region are the
next observational steps required to resolve the nature of this
puzzling object.

\section*{Acknowledgments}
We thank Thomas Erben and Yannick Mellier for providing the UH8K
$I$-band image of Abell 1942, and Alexandre Refregier for numerous
helpful discussions.  The Cambridge Infrared Survey Instrument is
available thanks to the generous support of Raymond and Beverly
Sackler.  MEG acknowledges support from the Canadian Cambridge Trust
and the Natural Sciences and Engineering Research Council of Canada.

\bsp

\label{lastpage}

\end{document}